\documentclass[%
 reprint,
superscriptaddress,
 amsmath,amssymb,
 aps,
floatfix,
]{revtex4-2}

\usepackage{graphicx}% Include figure files
\usepackage{xcolor}
\usepackage{dcolumn}% Align table columns on decimal point
\usepackage{bm}% bold math
\usepackage{soul}
\usepackage[caption=false,labelformat=simple]{subfig}

\usepackage{physics}
\usepackage{siunitx}

\usepackage{hyperref}% add hypertext capabilities
\usepackage{amsthm}
\newtheoremstyle{query}%
{}{}%space above/below
{\color{red}}%body style
{}%heading indent
{\sffamily\bfseries}{:}{12pt}%heading style/punctuation/space after
{}% head spec
\theoremstyle{query}
\newtheorem{aq}{Author Query/Comment}

\newcommand{\baq}{\begin{aq}}%This just makes things easier
\newcommand{\eaq}{\end{aq}}

\begin{document}

\preprint{APS/123-QED}
\title{Stimulated Laser Cooling Using Microfabrication}

\author{Chao Li}
\altaffiliation[Now at ]{Research Laboratory of Electronics, Massachusetts Institute of Technology, Massachusetts 02139, USA}
\author{Xiao Chai}
\author{Linzhao Zhuo}
\author{Bochao Wei}
\affiliation{School of Physics, Georgia Institute of Technology, 837 State Street, Atlanta, Georgia 30332, USA}
\author{\\Ardalan Lotfi}
\author{Farrokh Ayazi}
\affiliation{School of Electrical and Computer Engineering, Georgia Institute of Technology, 777 Atlantic Drive NW, Atlanta, Georgia 30332, USA}
\author{Chandra Raman}
\email{craman@gatech.edu}
\affiliation{School of Physics, Georgia Institute of Technology, 837 State Street, Atlanta, Georgia 30332, USA}

\date{\today}% It is always \today, today,

\begin{abstract}
We have achieved stimulated laser cooling of thermal rubidium atomic beams on a silicon chip.  Following pre-collimation via a silicon microchannel array, we perform beam brightening via a blue-detuned optical molasses.  Owing to the small size of the chip elements, we require only 8~mW, or nine times lower power than earlier free-space experiments on cesium [Aspect \emph{et al.}, Phys. Rev. Lett. \textbf{57}, 1688 (1986)].  Silicon micromirrors are fabricated and hand-assembled to precisely overlap a strong elliptical standing wave with a sheet-shaped atomic density distribution, with dimensions chosen precisely to match these.  We reduce the transverse velocity spread to below 1~m/s within a total travel distance of 4.5~mm on a silicon substrate.  We use Doppler-sensitive two-photon Raman spectroscopy to characterize the cooling.  In contrast to time-of-flight methods utilized previously, this approach requires a much shorter apparatus to achieve similar resolution.  This hybrid of passive and active collimation paves the way toward the construction of full-fledged atomic instruments, such as atomic beam clocks and gyroscopes, entirely on-chip through batch-fabricated processes.

\end{abstract}

\maketitle
%\tableofcontents

\begin{figure*}[!t]
%\centering %The revtex package documentation advises against the use of \centering (or \begin{center} and \end{center}) commands in figure and table environments
         \includegraphics[width=0.75\textwidth]{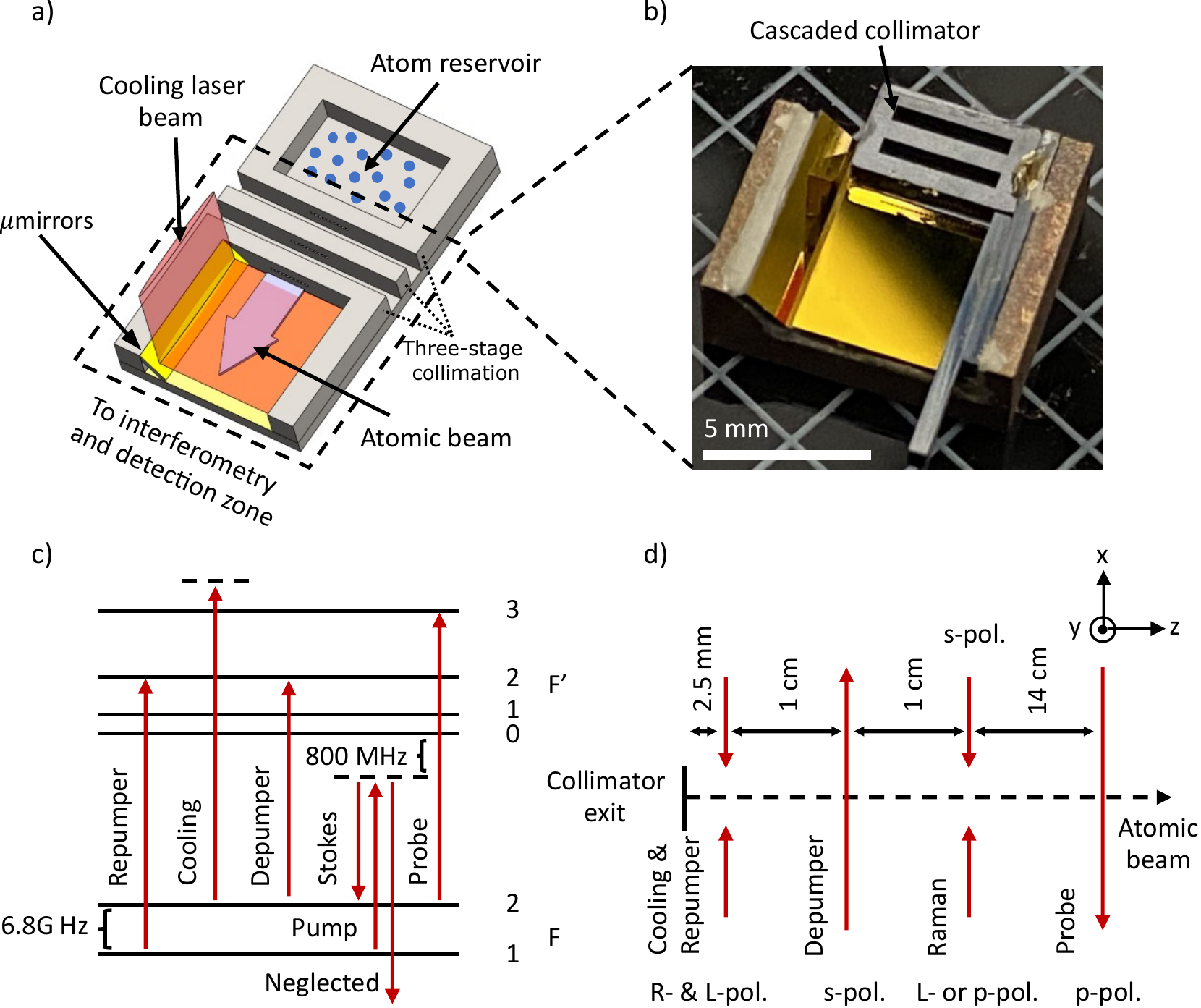}
         \caption{Overview of the transverse laser cooling experiment and its benchmarking using Doppler-sensitive Raman velocimetry. (a)  Conceptualized device containing an atom reservoir, multiple stages of a microcapillary array, and micromirrors. 
         Microchannels direct a continuous stream of atoms to the cooling region with pinpoint accuracy, where micromirrors maneuver laser beams to form a strong standing wave, compressing transverse velocity to the $<$1~m/s range within a distance of a few millimeters. 
         (b) Photograph of a manual assembly, containing a microfabricated cascaded collimator made of silicon, alumina-protected gold mirrors, and an EDM copper framework mounting the collimator and laser routing optics. (c) Energy-level diagram of the $^{87}$Rb D$_2$ transitions along with light fields, shown in dark red, used for laser cooling, optical pumping, two-photon Raman transition, and fluorescence measurement. (d) Top view of the experimental configuration, showing the atomic beam (dashed line), going through a sequence of light fields, with their relative spacing marked. ``R/L-pol.'' is right/left-handed circular polarization with respect to the direction of propagation. ``s/p-pol.'' is linear polarization perpendicular/parallel to the plane of incidence $x\text- O \text- z$.}
         \label{fig:concept}
\end{figure*}

%%%%%%%%%%%%%%%%%%%%%
\section{Introduction}
Atomic instruments such as precision gravimeters \cite{menoret2018gravity} and clocks \cite{knappe2004microfabricated} are increasingly making their way to the marketplace.  However, current manufacturing methods rely heavily on hand assembly, which drives up  cost and lowers  yield.  Much research effort has been focused on  laser-cooled atoms at the chip scale \cite{keil2016fifteen}, where there is potential for the use of batch fabrication methods that can address some of these challenges.  However, the power of automated fabrication has not yet been harnessed for atomic instruments.  Existing devices largely rely on collection of atoms within a three-dimensional (3D) vapor magneto-optical trap (MOT) before they are transferred to the chip.  This approach has been the workhorse of laboratory-scale experiments owing to its great flexibility, but it needs to be rethought for practical sensor deployment.  For example, it places an undue burden on manufacturing techniques to produce ultrahigh vacuum levels simply to load the MOT to a sufficient density.  It also necessitates large (centimeter-sized) cell geometries that are not compatible with semiconductor manufacturing and places atomic sensors in a pulsed, rather than a continuous, mode of operation, where there are unwanted dead times that must be accounted for.

Passively collimated atomic beams represent an alternative approach for atomic sensors that do not suffer from the above limitations.  Many applications such as atom interferometry \cite{berman1997atom}, guiding \cite{takekoshi2007optical}, and cavity quantum electrodynamics (QED) \cite{thompson1992observation,terraciano2009photon} have relied on atomic beam techniques.  Transverse laser cooling can substantially increase atomic beam brightness for such experiments \cite{slowe2005high}.  Unfortunately, many of the miniaturization gains from an on-chip architecture evaporate when  weak spontaneous force laser cooling is used, since this  requires centimeter distances to take effect.  By contrast, stimulated forces can be much stronger and can reduce the slowing distance, although, in free-space experiments, this type of laser cooling has the disadvantage of typically requiring laser power in excess of 100~mW.

In this work, we demonstrate stimulated laser cooling of a thermal rubidium atomic beam on a chip within an operating distance of only 4.5~mm.  In contrast to earlier work performed on free-space atomic beams \cite{aspect1986cooling}, we show that the reduction of volume at the chip scale results in a substantial reduction in laser power to only 8~mW without the need for multipass geometries and/or optical cavities, which could be added in a further step \cite{tanner1988atomic,hoogerland1996bright}.  The velocity spread is reduced by a factor of five while still cooling more than $10^{10}$ atoms/s.  These results pave the way toward on-chip guiding and similar applications needing a higher degree of collimation such as that available from a 2D MOT.  
Our setup, shown in Fig.~\ref{fig:concept}, features silicon micromirrors bonded together with silicon atomic beam collimators \cite{li2019cascaded,li2020robust} and a silicon substrate on a copper frame to achieve an overall length of only 8 mm.
Moreover, we have characterized the cooling in detail using an elegant and compact method based on Doppler-sensitive two-photon Raman spectroscopy.  By contrast, traditional atomic beam experiments have used time-of-flight characterization over meter distances to achieve  a similar velocity resolution \cite{aspect1986cooling,chen1992adiabatic}.  A comparison between the two methods is presented.

%%%%%%%%%%%%%%%%%%%%%
\section{Experiment}

In Fig.~\ref{fig:concept}(a), we present a future chip-scale atomic beam device that can  take full advantage of micro-electromechanical systems (MEMS) engineering and vacuum packaging technology \cite{kitching2018chip,li2019cascaded}. 
For example, deep reactive ion etching (DRIE) of silicon can define functional regions for beam formation,  ballistic propagation, and interferometry. Wafer bonding can hermetically seal  silicon functional and glass capping layers, offering optical access.
Using these methods, typical vapor-cell atomic sensors adopt vertical integration in which incident laser beams go through a stack of device layers sandwiching atoms. 

To reduce laser power consumption, we implemented cooling along an axis within the plane of the wafer,  with the smaller cross-sectional area of the planar atomic beam sheet allowing us to focus more tightly.   This allowed us to use slanted micromirrors.  Such mirrors can be 
realized using a variety of microfabrication protocols \cite{ahn2021large,rola2014silicon,brockmeier2012surface,Chutani2014}, some of which have found  application in making chip-scale pyramidal MOTs \cite{trupke2006pyramidal}, vapor-cell clocks \cite{nishino2019reflection,chutani2015laser}, and NMR gyro \cite{noor2017status},but not previously in the context of laser cooling for thermal atomic beams. In this work, a sub-millimeter proximity between the mirrors and the atomic beam was key to the device operation.  Therefore, a significant effort went toward ensuring mirror surfaces were protected against alkali attack, as we describe below.
 
In the following, we also present details of how we manually assembled a chip-scale laser cooling platform. 
%We describe a compact oven design for atomic beam production.
Then, we describe our laser systems and relevant optical setup for atomic state preparation, laser cooling, Raman velocimetry, and detection. 

\subsection{Laser cooling assembly}  
The silicon collimator shown in Fig.~\ref{fig:concept}(b) has dimensions height $\times$ width $\times$ length = $1\;\mathrm{mm} \times 5\;\mathrm{mm} \times 3\;\mathrm{mm}$. There are 20 thin microcapillaries, each with a square cross-section and of 100~$\mu$m diameter. Two 600~$\mu$m gaps were machined from the top, halfway through, using a femtosecond laser cutter, partially exposing the microchannels to free space. The characteristics of these silicon collimators, such as their collimating performance, long-term robustness, and  compatibility with different types of atom sources, including pure rubidium and dispenser, have been thoroughly studied in our previous work \cite{li2019cascaded,li2020robust,wei2022collimated}.

The vertical, bottom, and slanted mirrors have dimensions $1\;\mathrm{mm} \times 8\;\mathrm{mm}$, $5.5\;\mathrm{mm} \times 8\;\mathrm{mm}$, and $2\;\mathrm{mm} \times 6.5\;\mathrm{mm}$, respectively. Here, we briefly summarize the fabrication recipe for making these protected gold mirrors that are resistant to alkali attack.
An electron beam evaporation technique was used to form the mirror coating on a double-side-polished 500~$\mu$m thick silicon substrate. These layers included 20~nm chromium as an adhesive layer, 100~nm gold for increasing reflectivity, and 100~nm alumina (Al$_2$O$_3$) as a protective layer \cite{slowe2005high}. The coated substrate was diced to the desired dimensions, and xenon fluoride gas was used to clean the mirror pieces by removing silicon particles generated during the dicing step.

All components were manually glued to an electrical-discharge-machined (EDM) copper framework using vacuum-compatible epoxies (EPO-TEK 353ND and Master Bond Supreme 18TC). Unfortunately,  the edges of the vertical mirror were damaged during the gluing process, and therefore a replacement mirror was overlaid on top of the damaged mirror [it appears as the extruded piece in Fig.~\ref{fig:concept}(b)]. 
In addition, an extra piece of thin silicon with approximate dimensions $8\;\mathrm{mm}\times14\;\mathrm{mm}$ was mounted vertically, immediately above the third stage of the cascaded collimator  (not shown). This extra piece was to prevent off-axis atomic vapor from entering the atomic beam region downstream, as in our previous work \cite{li2019cascaded,li2020robust}. 
The laser cooling assembly, including the copper framework, has a total volume of $\sim$0.25~cm$^3$. The silicon collimator and mirrors occupy about only half of it. This miniature assembly was then attached to a conventional rubidium oven, as in our previous work \cite{li2019cascaded,li2020robust}. 
A detailed description of a compact oven design for atomic beam production is given in the Appendix.

\subsection{Laser systems and atomic transitions} 
We now describe the light fields through which an atomic beam sequentially passed according to the energy level diagram in Fig.~\ref{fig:concept}(c) and and the schematic of the experimental configuration in Fig.~\ref{fig:concept}(d). The diameters and powers of the various laser beams  are summarized in Table~\ref{tab:table2}. We used two separate diode laser systems (DLC DL PRO 780 S) for the repumper and the cooling beams, respectively. The repumper was on-resonance with the $F=1$ to $F'=2$ transition. The stimulated transverse cooling beam was blue-detuned from the $F=2$ to $F'=3$ transition by $\delta_c/2\pi$~MHz, where $\delta_c/2\pi$ could be tuned in the range from 40~MHz to 95~MHz, corresponding to the range from $6.6\Gamma$ to $16\Gamma$. Here, $\Gamma=2\pi \times 6$~MHz is the natural linewidth of the $^{87}$Rb D$_2$ transition. By assigning a red detuning of these values  the sign of the stimulated forces could be flipped,  causing heating of the atoms. 
Using a tapered amplifier (MOGLabs MOA003), we were able to provide enough power to create an intense standing wave for the stimulated transverse cooling over the range of laser powers used in this work.
A separate repumper was merged with the cooling beam using a fiber polarization combiner (Thorlabs PFC780A). We set the cooling beam output  to $\sigma^+$ to achieve optimized laser cooling for the stretched states $\ket{F=2, m_F=2}$ and $\ket{F=3, m_F=3}$. For the laser cooling process, the intense light field was sufficient to clearly define the quantization axis along $+\hat{x}$ for $m_F$. For example, 1~mW of laser power averaged over the small cooling beam size led to a Rabi frequency of $6\Gamma$, which was much larger than the Larmor frequency of $^{87}$Rb, which is sub-MHz under the Earth's magnetic field.

We performed Raman spectroscopy using a separate distributed Bragg reflector laser (Photodigm Mercury TOSA DBR), red-detuned by 800~MHz from the $F=1$ to $F'=0$ transition. This beam passed through a fiber-coupled electro-optic modulator (EOM, PM-0K5-10-PFA-PFA-780), with a fed-in microwave power (from a Windfreak SynthHD PRO v2) of 10.6~dBm and frequency near $\omega_{HFS}/2\pi$, which is the hyperfine splitting of the $F=1, 2$ ground states. We set the EOM power to produce roughly equal zeroth- and first-order sidebands, which were used as the pump and Stokes beams for the Doppler-insensitive spectroscopy.  We scanned their difference frequency via the microwave source, with the step time and step frequency set as 1~ms and 0.02~MHz. Therefore, it took the microwave source 5~s to fully sweep $\pm$50~MHz over the hyperfine resonance.

\begin{table}[!t]
\caption{Diameters and powers of laser beams in Fig.~\ref{fig:concept}.\label{tab:table2}}
\begin{ruledtabular}
\begin{tabular}{lcdd}
Laser beam & $2w_z$ (mm)\footnotemark[1] & \multicolumn{1}{c}{$2w_y$ (mm)\footnotemark[1]} & \multicolumn{1}{c}{Power (mW)}  \\
\hline
Repumper & 4.5 & 0.21 & \multicolumn{1}{c}{6--8}  \\
Cooling & 4.5 & 0.21 & \multicolumn{1}{c}{3.0--61} \\Depumper &  2.0 &  2.0 &  8 \\
Raman Stokes & 1.0 & 1.0 & 0.3 \\
Raman pump (Co.)\footnotemark[2] & 1.0 & 1.0 & 0.3  \\
Raman pump (C.)\footnotemark[3] & 0.8 & 0.8 & 2.0 \\
Probe  & 5.8 & 5.8 & \multicolumn{1}{c}{5--7} \\
\end{tabular}
\end{ruledtabular}
\footnotetext[1]{$w_z$ and $w_z$ are the $1/e^2$ Gaussian beam waists.}
\footnotetext[2]{Co-propagating component for Doppler-insensitive cases.}
\footnotetext[3]{Counter-propagating component for Doppler-sensitive cases.}
\end{table}

\begin{figure*}[!t]
         \subfloat[]{\includegraphics[width=0.8\textwidth]{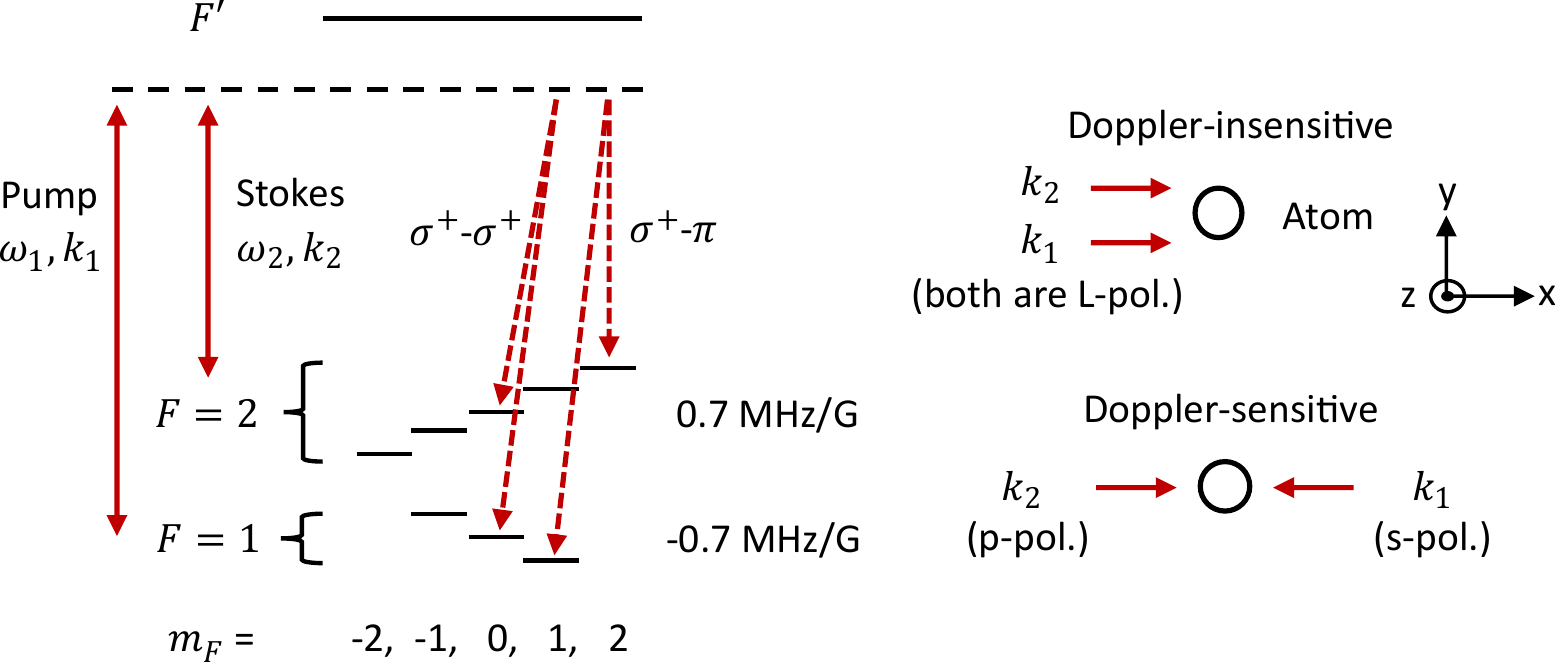}
         \label{fig:zeeman_sublevels}}
 \\
         \subfloat[]{\includegraphics[width=0.4\textwidth]{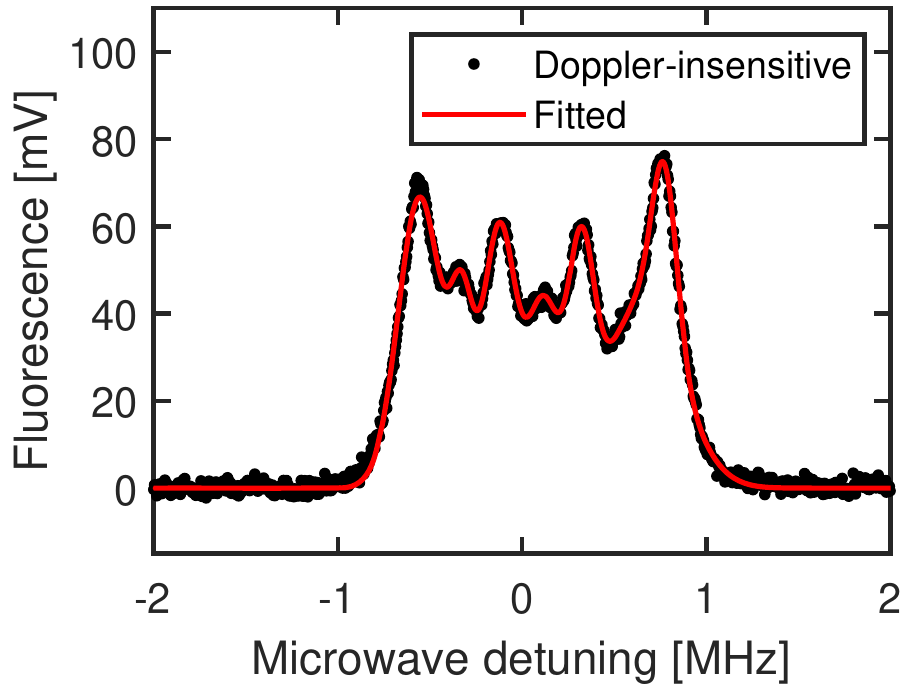}
         \label{fig:dop_insensitive}}
         \subfloat[]{\includegraphics[width=0.4\textwidth]{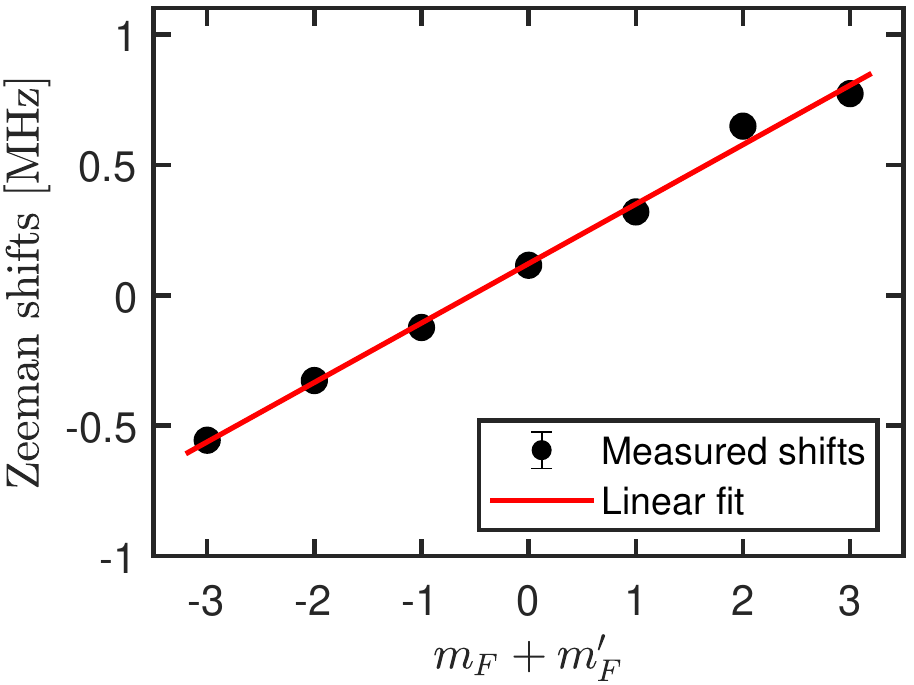}
         \label{fig:zeeman}}   
     \caption{Working principle of  Raman velocimetry. (a) Relevant energy levels of the $^{87}$Rb D$_{2}$ transition along with co-propagating/counter-propagating light fields driving the Doppler-insensitive/sensitive Raman transition. In the left panel, shown are two representative transitions with $\Delta m_F^o=0$ or 1, which is the overall change of $m_F$ after a two-photon transition. Here, polarization states are defined with respect to a quantization axis along the earth field (see text for discussion).  (b) Measured Doppler-insensitive Raman spectra containing transit-time, power and Zeeman broadening. (c) Zeeman shifts extracted for magnetic sublevels. $m_F+m'_F=0$ and $\pm2$ are for $\Delta m_F^o=0$ transitions. $m_F+m'_F=\pm1$ and $\pm3$ are for $\Delta m_F^o=\pm1$ transitions. }
     \label{fig:technique}
\end{figure*}
\subsection{Fluorescence detection}\label{section:fluo_det}
The passive collimation and active laser cooling took 8~mm of the propagation distance in total, as shown in Fig.~\ref{fig:concept}(b). Preparing atomic states, Raman velocimetry, and probing the fluorescence took another 16~cm. For the Doppler-sensitive configuration, the Raman process transferred atoms pre-depumped in the $F=1$ state back to $F=2$ for those satisfying $\delta_r = 2kv_x$ within the narrow linewidth of the Raman transition, where $\delta_r$ is the two photon detuning. The fluorescence of the $F=2$ state was detected downstream on a dark background without suffering from scattered light from other laser beams, as illustrated in  Fig.~\ref{fig:concept}(d). 

\begin{figure*}[!t]
         \subfloat[]{\includegraphics[width=0.4\textwidth]{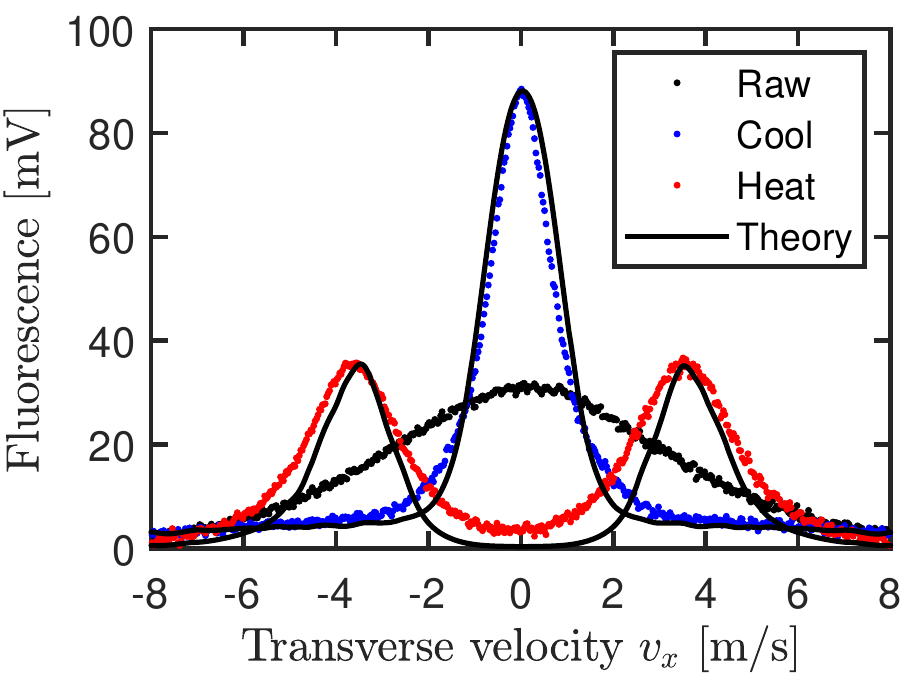}
         \label{fig:raw_cool_heat}}
          \subfloat[]{\includegraphics[width=0.6\textwidth]{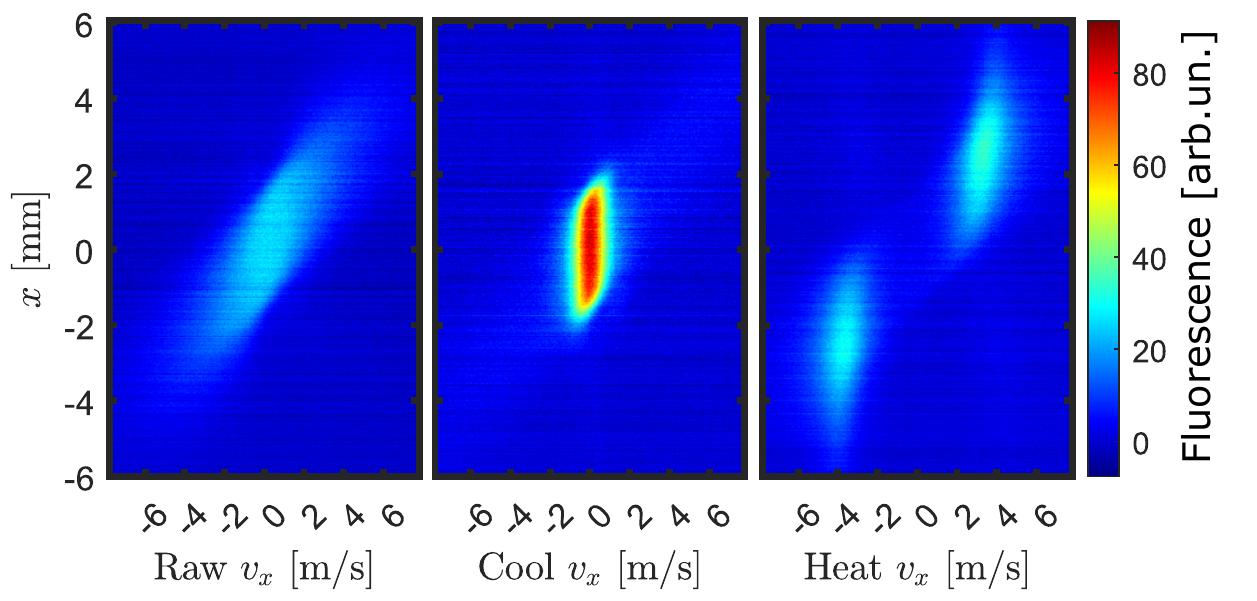}
         \label{fig:velocity_map}}
     \caption{Raman velocimetry for thermal atomic beams perturbed by an intense laser standing wave. 
     % (HWHM = 3.6$\pm$0.5 m/s)
     % (HWHM = 0.84$\pm$0.05 m/s)
     (a) Measured fluorescence vs two-photon detuning $\delta_r=2kv_x$ of the Doppler-sensitive Raman transition, where $k$ is the wavevector of the $^{87}$Rb D$_2$ transition and $v_x$ is the transverse velocity of the atoms. Black dots (HWHM = 4~m/s) are for the cooling beam off; blue dots (HWHM = 0.8~m/s) are for the cooling beam on with positive detuning ($\delta_c/2\pi = +40$~MHz), and red dots are for the cooling beam on with negative detuning ($\delta_c/2\pi = -40$~MHz). The dashed curves are theoretical predictions given by the Fokker–Planck equation.
     (b) Camera footage showing the changes in both fluorescence intensity and position as the two-photon detuning $\delta_r$ is scanned.}
     \label{fig:velocimetry}
\end{figure*}

The fluorescence spectra and images presented in the following subsections were collected by a set of 2~in. ($\sim$5~cm) lenses 9~cm above the probe region, with a collection efficiency of about 1.3\%.
The output of the photodiode  (Thorlabs DET100A2) was converted to voltage by a current amplifier  (DL Instruments 1211) with a gain setting of $10^9$~V/A and a claimed minimum rise time of 0.25~ms.
We used a coefficient of $6\times10^{10}$~atoms\,s$^{-1}$\,V$^{-1}$ for converting the measured fluorescence voltage to atomic flux, counting both $^{87}$Rb and $^{85}$Rb. We performed data acquisition  using an oscilloscope with a 10~kS/s sampling rate at  12-bit resolution (PicoScope 5242D). We averaged 40 neighboring  data points when a better comparison was needed in scenarios such as that shown in Fig.~\ref{fig:raw_cool_heat}.

We  also determined the correlation between the transverse velocity distribution and the spatial spread of the atomic beam in the probe region by directly replacing the photodiode by a camera with $2464\,(x) \times 2056\,(z)$ pixels with unit-cell size  3.45~$\mu$m (Imaging Source DMK 33UX264). The imaging system had a magnification  $M=0.7$, which was used to convert the images of atoms to their spatial locations. However, the fluorescence imaging method operated more slowly than the photocurrent detection method, because the exposure time was set to be 1/10~s or 1/15~s to provide sufficient sensitivity. The former setting was for recording the raw and heated spectra, and the latter  for the laser-cooled case. We also reduced the microwave source scan range to be $\pm20$~MHz with 0.1~MHz step size per camera exposure time to accommodate the slow response. These camera footages are given in Fig.~\ref{fig:velocity_map} and Fig.~\ref{fig:selected_cam_frm}.

\subsection{Raman velocimetry and its resolution}
As shown in Fig.~\ref{fig:zeeman_sublevels}, two light fields coherently and adiabatically transferred atoms from the ground state $\ket{F=1}$ to $\ket{F=2}$ via  stimulated Raman transition. 
We used co-propagating circularly polarized beams for Doppler-insensitive transitions and counter-propagating lin$\perp$lin for Doppler-sensitive ones.
To achieve the latter, we sent an extra portion of the DBR laser power from the opposite direction.

We measured the Doppler-insensitive Raman transition under weak excitation without canceling the Earth's magnetic field and other stray fields. 
The atomic beam  traveled along $+\hat{z}$, roughly pointing to 32$^\circ$NE in the laboratory with geographic coordinates (33$^\circ$N, 84$^\circ$W). 
Therefore, the $k$-vectors of the Raman beams were not aligned to the direction of the magnetic field, and  all three projected $\sigma^{+}$, $\sigma^{-}$, and $\pi$ components were present after decomposition with respect to a quantization axis defined by the direction of the Earth's field.
Each Raman beam could thus drive all allowed one-photon transitions ruled by $\Delta m_{F}=-1,0,1$ \cite{desavage2013controlling,desavage2011raman}. 
There were then nine possible types of transitions in total, given by two photons labeled with an arbitrary pair of polarization states. The 
Zeeman shifts associated with those transitions were equal to $(m_F+m'_{F})\times0.7\,\mathrm{MHz/G} \times B$, considering an overall change in the magnetic quantum number following $\Delta m_F^o = m'_{F}-m_F = \pm0, \pm1, \pm2$, with the $\pm2$ transitions being neglected owing to their much smaller transition strengths.

Fig.~\ref{fig:dop_insensitive} shows the measured Doppler-insensitive Raman spectra zoomed in the $\pm$2~MHz range from a full sweep of $\pm$50~MHz as we scanned the microwave source.
We identify seven peaks from a fit of the data to seven Gaussian functions according to selection rules \cite{desavage2013controlling}, with the sixth less prominent buried under two neighboring protrusions.
The peak centers are plotted vs $m_F+m'_{F}$ in Fig.~\ref{fig:zeeman}. A linear fit reveals the existence of  $0.3$~G magnetic fields.
The clock transition was shifted by 0.12 MHz, i.e. greater than the theoretically estimated AC Stark shift of $<$ 0.02 MHz.  We attribute this to small errors in the microwave scan much smaller than our transit time-broadening of $\sim$0.4~MHz (FWHM).
In addition, the fact that these seven peaks  overlap with each other also implies a transit-time broadening of $\sim$0.4~MHz for a 1~mm diameter Raman beam.
According to the measured spectral half width at half maximum  (HWHM) $\sim$1~MHz shown in Fig.~\ref{fig:dop_insensitive}, we can claim that the Doppler resolution of our Raman velocimetry is $\sim$0.4~m/s. 
Canceling the earth's field would improve this resolution by another factor of five, ultimately limited by the transit-time broadening. 

We performed Doppler-sensitive Raman velocimetry at higher laser power (see Table \ref{tab:table2}) to (i) enhance the signal-to-noise ratio and (ii) enable the Raman process on a faster time scale compared with Larmor precession. This allows us to discuss the Raman process with respect to a quantization axis along the propagation direction of the Raman beam.
The effective Rabi frequency governing a coherent population transfer from $\ket{F=1}$ to $\ket{F=2}$ is simply proportional to $\sum(E_{p}\bra{F=1, m_F} d \ket{F', m_{F'}} E_{s}\bra{F=2, m_F} d \ket{F', m_{F'}})$, counting all allowed transitions, where $E_{p}$ and $E_{s}$ are the pump and Stokes electric fields, and $\bra{F, m_F} d \ket{F', m_{F'}}$ are their associated transition dipole matrix elements between the ground state $\ket{F, m_F}$ and the excited state $\ket{F', m_{F'}}$ \cite{steck2021rubidium}. 

We then decompose the pump and Stokes fields into circular polarization bases, according to $\hat{\bm{z}} = (1/\sqrt{2})(-\hat{\bm{e}}_+ + \hat{\bm{e}}_-)$ and $\hat{\bm{y}} = (i/\sqrt{2})(\hat{\bm{e}}_+ + \hat{\bm{e}}_-)$, where $\hat{\bm{e}}_{+/-}$ drives the $\sigma_{+/-}$ transition. 
Our Doppler-sensitive laser setup composes two counter-propagating laser beams, one of the beam is polarized along z, with only the Raman pump frequency component. The other beam is polarized along y and EOM modulated to have both the Raman Stokes and pump frequency components. Due to quantum destructive interference, the co-propagating Raman pair can only weakly drive the Raman transition, while the counter-propagating Raman pair has much larger effective Rabi frequency. Therefore the effect of the 0th EOM side band is ignored and this configuration is effectively Doppler-sensitive.

%%%%%%%%%%%%%%%%%%%%%
\section{Results}\label{sec:results}

Hot-wire ionizing or fluorescence imaging techniques deduce the span of transverse velocities from the spatial spread of atoms \cite{aspect1986cooling,chen1992adiabatic,mitra2020direct}.
Assuming all atoms emerge from a source with negligible size and an atom is probed in the far field at location $z$, its transverse velocity can be estimated according to $v_x=vx/z$, where $x$ is the offset of an atom from the central axis, and $v$ is the longitudinal velocity. 
Therefore, these techniques can not be applied to a dense array of atomic beams with a few millimeter width, especially when the beam propagation distance is limited within $\sim$centimeter, as in chip-scale atomic beam devices.
In contrast to probing the atoms spatially,  two-photon Raman velocimetry can directly resolve the transverse velocity in the frequency domain, where the transit-time broadening, given the finite size of laser beams driving the Raman transition, sets the limit on Doppler resolution.

Fig.~\ref{fig:velocimetry} shows our principal data. An incident power of only 7.8~mW creates an intense standing wave with a maximum Rabi frequency of $72\Gamma$ under laser detunings of $\pm6.6\Gamma$, comparable to that of Ref. \cite{aspect1986cooling}. 
In Fig.~\ref{fig:raw_cool_heat}, we plot the measured fluorescence vs the microwave detuning under three scenarios. All data sets presented here and in the rest of the paper have subtracted background measured at $v_x=20$~m/s.
The raw spectrum shows an original transverse velocity distribution without  perturbation by the laser standing wave. The measured HWHM of 4~m/s is in very good agreement with our previous results for an identical cascaded collimator that was characterized via a Doppler-sensitive one-photon cycling transition \cite{li2019cascaded}. In our previous study, artifacts already appeared in the wings of the deconvolved velocity distribution because the intrinsic power-broadened natural linewidth contributed significantly to the measured spectral width.
In this work, as can be seen from  Fig.~\ref{fig:dop_insensitive}, the two-photon Raman transition possesses a HWHM linewidth $<$1~MHz,  which is much smaller than the Doppler HWHM of our atomic beams $\sim$10~MHz. Therefore, a deconvolution to retrieve the transverse velocity distribution is unnecessary for the quick comparison given here. The cooled spectrum shows that the width of the velocity distribution immediately shrinks by a factor of five once the laser cooling is turned on. 
The ratio of measured cooled and raw spectra at near-zero $v_x$ implies that the brightness of the atomic beam  has been enhanced by a factor of three. 
The heated spectrum reveals that the sign of the stimulated forces has been flipped, and atoms are pushed toward the shoulder protruding around $\pm$3.5~m/s.

The theoretically predicted fluorescence in Fig.~\ref{fig:raw_cool_heat} is calculated by solving the following Fokker--Planck equation (see, e.g., \cite{berg1992momentum,chen1993evolution,drewsen1993intensifying,molmer1994optimum}):
\begin{equation}
    \pdv{p}{t} =-\frac{1}{m} \pdv{v_x}(Fp) + \frac{1}{m^2}
    \pdv[2]{v_x}(Dp),
\end{equation}
where $p(v_x,t)$ is the transverse velocity distribution, $m$ is the atomic mass, and $F(v_x)$ and $D(v_x)$ are velocity-dependent force and diffusion coefficients, respectively. Given the Rabi frequency $\Omega$ and detuning $\Delta$, $F(v_x)$ and $D(v_x)$ are evaluated using a continued fraction method \cite{minogin1979resonant,berg1992momentum}. For simplicity, we ignore the spatial dependence of the laser beam intensity and approximate the Rabi frequency as its average value $\Omega=36\Gamma$ \cite{minogin1979resonant}. The evolution time is 12~$\mu$s, as given by $t_e = 2 w_z / \bar{v}_B$, where $2w_z=4.5$~mm is the beam diameter and $\bar{v}_B=\SI{372}{\metre/\second}$ is the atomic beam mean velocity along the longitudinal direction given a nozzle temperature of 135 C \cite{ramsey1985molecular}. The initial condition is deconvolved from the raw experimental data in Fig.~\ref{fig:raw_cool_heat}.

In Fig.~\ref{fig:velocity_map}, we present  camera footage of the fluorescence intensity vs $v_x$, directly converted from time according to the parameters we used for the linear sweep of the microwave source.
Given each velocity bin, the fluorescence intensity presented along $x$ is an average over $z$ for the region of interest (ROI) extracted from a single video frame, where $x$ and $z$ are the propagation directions of the probe beam and the atomic beam, respectively (see Appendix and Sec. \ref{section:fluo_det}).
These data not only independently validate the fluorescence spectral information, i.e., a brightness enhancement by a factor of three, as given in Fig.~\ref{fig:raw_cool_heat}, but also reveal the velocity-dependent spatial vibration of the atoms in a similar manner to recent laser cooling work on molecular beams \cite{mitra2020direct}, although the results of the latter do not directly contain frequency-domain information.

\begin{figure*}[!t]
          \subfloat[]{\includegraphics[width=0.45\textwidth]{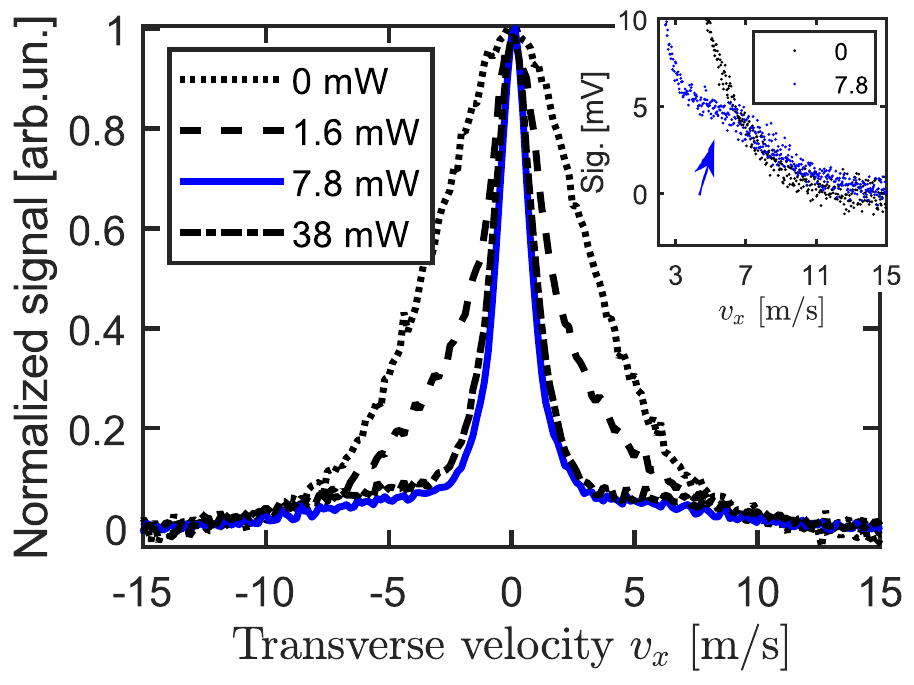}
         \label{fig:spectra_vs_power}}
          \subfloat[]{\includegraphics[width=0.515\textwidth]{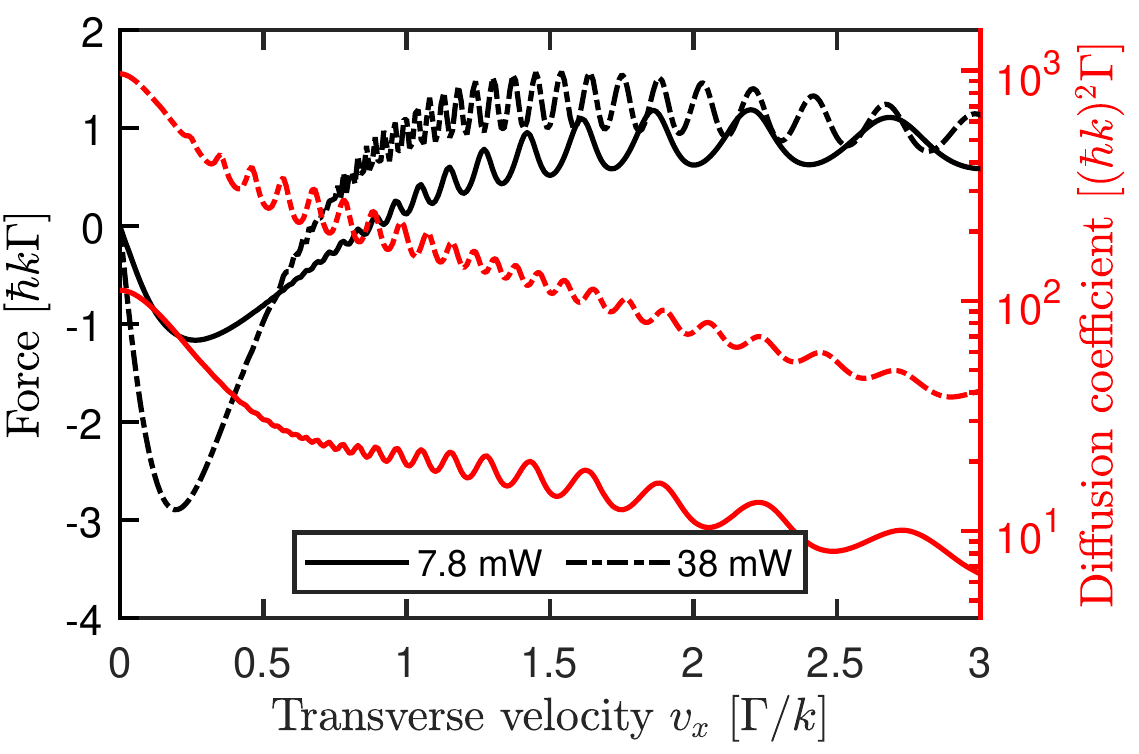}
          \label{fig:force_diffu}}\\
          \subfloat[]{\includegraphics[width=0.5\textwidth]{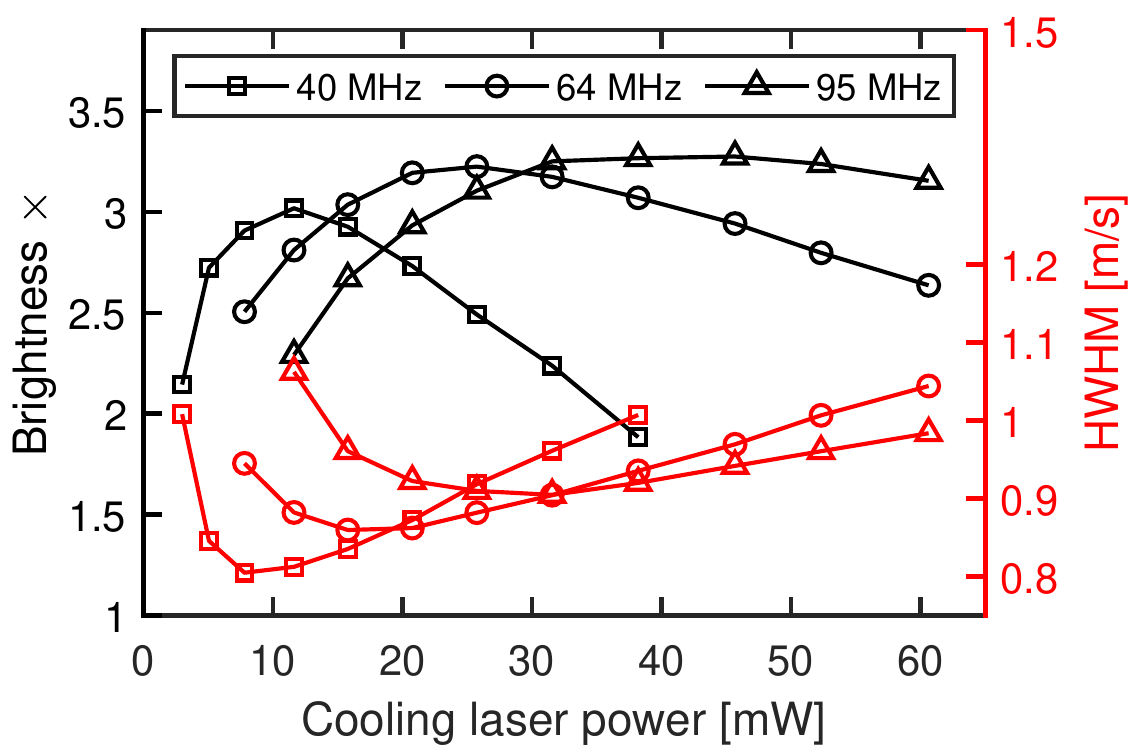}
         \label{fig:height_width}}     
        \caption{Power and detuning dependence of laser cooling. (a) Fluorescence signal normalized to peak value vs power shows the progression of laser cooling. The inset zooms into the velocity tail of the raw and best-cooled spectra before normalization.  (b) Corresponding dipole forces and diffusion coefficients  computed for the best-cooled and overshoot cases  in (a). Solid lines are for lower power and dashed lines are for higher power. The force/diffusion profile is an odd/even function, and so the negative-velocity region is not shown. In (a) and (b), $\delta_c/2\pi = +40$~MHz. (c) Brightness gain (black curve) and the corresponding spectral width (red curve). Square, circular, and triangular circles are experimental data collected at blue detunings of 40~MHz, 64~MHz, and 95~MHz, respectively.}
        \label{fig:optimization}
\end{figure*}

Let us now compare our experiment with a previous free-space experiment on Cs thermal atomic beams \cite{aspect1986cooling}. The previous experiment required more than 10~cm to  achieve a combination of passive and active collimation, which consumed 70~mW of laser power. 
Our experiment has successfully achieved a similar protocol on a sub-centimeter chip-scale platform consuming nine times less laser power. 
In addition, microfabricated collimators and mirrors are flush against each other during manual assembly. This naturally enables  precise alignment between the atomic beam and the laser standing wave in the cooling zone. 
Alignment uncertainty is measured to be $<1^{\circ}$ under an optical microscope after we glue the collimator and mirrors together.
Any significant misalignment would be manifested as an asymmetry between two shoulder peaks \cite{aspect1986cooling} given a spectral measurement like that in Fig.~\ref{fig:raw_cool_heat}, supposing that perfect orthogonality is achieved for all other laser--atom-beam intersections. 
In the third panel of Fig.~\ref{fig:velocity_map}, we note the existence of a slight asymmetry in fluorescence intensity  for atoms at $\sim\!\pm4$ m/s. 
This is mainly attributable to the imperfect orthogonality between the probe laser beam and the atomic beam for this particular data set.

Another unique advantage results from the 1D nature of an atomic beam array and the precision match between the laser mode and the vertical size of the atomic beams close to the collimator exit.
The laser cooling power required typically increases as an atomic beam gets larger. However, power consumption can stay constant as we pack more channels into this 1D microcapillary array, within the Rayleigh range for a given laser beam size, because one laser standing wave cools them all.
Compared with spatial probing methods, our Raman velocimetry diagnostic method does not require a propagation distance of meters, which was previously essential, to reveal the divergence angle of a well-collimated atomic beam. In principle, all experimental state preparation, Raman velocimetry, and detection sequences could be done within another sub-centimeter propagation distance of the atomic beam after the laser cooling stage by careful management of the scattered light. This is indispensable for benchmarking the performance of future fully integrated chip-scale atomic beam devices, considering the limited transit length on a chip.

In the following, we use Raman velocimetry to study the dependence of laser cooling on laser intensity and detuning and compare the data with theoretical predictions. Fig.~\ref{fig:spectra_vs_power} shows the progression of  cooling vs incident power. 
As the laser power goes up from 1.6~mW to 7.8~mW, stimulated forces become larger, shortening the transverse velocity damping time from 30~$\mu$s to 5~$\mu$s. We estimate these numbers by computing the slope of the force profile close to zero transverse velocity.
At 1.6~mW, the evolution time $t_e=12$~$\mu$s is not enough to accomplish the cooling process given a damping time of 30~$\mu$s, leading to the incomplete cooling shown as a dashed curve in Fig.~\ref{fig:spectra_vs_power}. 
At 7.8 mW the cooling is nearly complete.  However, one can distinguish fine details of the force profile from the wings of the spectrum.  The inset in the figure reveals a knee near 5 m/s.  This is near the region where the force flips its sign close to $\Gamma/k$, and agrees with the theoretical force profile shown in Fig.~\ref{fig:force_diffu}, where the solid black curve crosses zero force also at $\sim\!\Gamma/k$, defining the capture range for transverse velocities.
The measured spectra at 38~mW also suggest excessive laser power at a fixed detuning can cause the transverse temperature to increase due to larger dipole force fluctuations at a higher light intensity.

In addition to results at a blue detuning of 40~MHz, we replicate the same velocimetry measurement at $\delta_c/2\pi= 50$~MHz, 64~MHz, 80~MHz, and 95~MHz. Three representative data sets are shown in Fig.~\ref{fig:height_width}. 
This convenient approach allows us to determine a minimum HWHM and a maximum brightness enhancement at each detuning as we scan the cooling power across a broad range. Furthermore, increasing both detuning and power in a coordinated manner can further enhance the maximum brightness gain. We attribute this to a larger velocity capture range accumulating and cooling more atoms towards zero $v_x$ according to computed force profiles at larger detunings and higher intensities. An example of a force profile at a Rabi frequency of $1000\Gamma$ and a detuning of $200\Gamma$ can be found in Ref.~\cite{dalibard1985dressed}.

In summary, we have demonstrated stimulated laser cooling on a chip, reducing atomic transverse speeds to 80 cm/s within a small, sub-centimeter geometry that includes both the atomic beam collimator and cooling region. Only 8 mW were needed, substantially lower than the 10s of mW typically used in larger geometries typical of free-space cooling.  Our results point to multiple possibilities for on-chip atom manipulation, including beam brightening for Raman interferometry and atom guiding.  These could open up new avenues for atomic inertial sensing on chip.

\section*{Acknowledgments}
C.L. acknowledges Alexandra Crawford for reading and commenting on the manuscript, Anosh Daruwalla and Benoit Hamelin for providing earlier test samples.

\section*{Author contributions}
C.L. and C.R. conceived the experiments. 
C.L. designed the assemblies, conducted the experiments, and analyzed the data with experimental assistance from X.C., L.Z., and B.W.
C.L. wrote the manuscript with input from A.L., X.C., and C.R.
X.C. performed theoretical calculations.  
A.L. fabricated the micromirrors. 
C.R. and F.A. supervised the study. 

\appendix

\section{Camera footage processing}\label{section:camera}
\begin{figure}[htbp]
    \centering
    \includegraphics[width=0.35\textwidth]{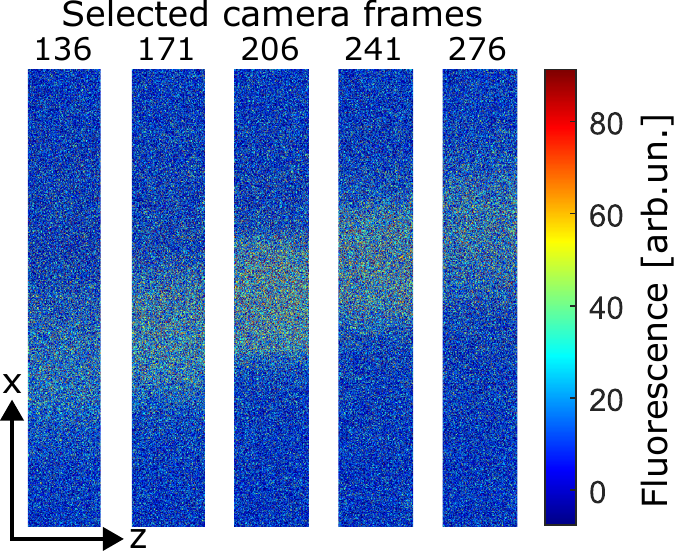}
    \caption{Fluorescence images after background subtraction of atoms in the $F=2$ state. The Doppler-sensitive two-photon Raman transition selects their transverse velocity. No laser cooling is in action for this data set, corresponding to the ``Raw'' case shown in Fig.~\ref{fig:velocimetry}(b). There are $2464\,(x) \times 400\,(z)$ pixels for the ROI of each frame.}
    \label{fig:selected_cam_frm}
\end{figure}

\section{Atomic beam oven design}\label{section:oven}
A 101 copper tube, of outer diameter 3/8~in. and length 4~in., serving as the main body of the oven, passes through a ring-shaped Ultem PEI spacer glued at the center of a bored blank KF40 flange. This  spacer thermally isolates the oven body from the supporting  flange. 
Another short pinched-off copper tube containing a glass ampoule of 200~mg rubidium is attached to the \emph{ex vacuo} end using an in-line Swagelok tube union. The capsule section can be replaced when all the rubidium is depleted.
On the \emph{in vacuo} end, there is a customized copper holder of diameter 5/8 in. with a rectangular opening at its center to accommodate and mate with the laser cooling assembly. 
This nozzle head is glued to the oven body with a 1~cm quartz tube in between, serving as another critical thermal break. This transparent quartz neck also offers access to measure the atomic vapor pressure inside the oven from absorption. 
The customized copper holder is externally threaded. A dual set of fine-gauge wires with insulating coatings is wrapped into the threaded grooves. 
One set of resistive wires is Nichrome; the other conductive set is copper. Without generating any bias magnetic fields, running DC through them maintains the nozzle head at a higher temperature than the oven body. It thus prevents rubidium from clogging the silicon collimator.

The KF flange surrounding the oven body contains double-pore ceramic electrical feedthroughs, which supply current to the resistive coil wrapped on the nozzle head and read the nozzle and neck temperatures with two 10~k$\Omega$ glass bead thermistors.
All joints are glued together using vacuum-compatible epoxies as mentioned above, with the caution that Master Bond Supreme 18TC is applied to regions requiring good thermal conductivity. 
Finally, the whole oven is attached to a $6\,\mathrm{in.}\times6\,\mathrm{in.}\times6\,\mathrm{in.}$ vacuum chamber, and the encapsulated rubidium ampoule can be cracked by compressing the copper tube once the vacuum pressure reaches $\sim\!10^{-5}$~Torr.
About half of the copper tube oven is outside the vacuum chamber and is heated by a rope heater. A PID controller and a thermocouple are used to keep the oven body temperature at $110\,^\circ$C. Temperature differences across the whole copper oven are $<15\,^\circ$C. A current of 0.16~A  is required to keep the nozzle head at $135\,^\circ$C, consuming $\sim$1~W of electrical power.
%($\sim\!\!15\,\mathrm{cm}\times 15\,\mathrm{cm}\times 15\,\mathrm{cm}$) 

\bibliography{references}% Produces the bibliography via BibTeX.

\end{document}